\documentstyle[11pt,epsf]{article}
\begin{document}
\thispagestyle{empty}
\rightline{UOSTP-00-103}
\rightline{{\tt hep-th/0006087}}

\

\def\tr{{\rm tr}\,} \newcommand{\beq}{\begin{equation}}
\newcommand{\eeq}{\end{equation}} \newcommand{\beqn}{\begin{eqnarray}}
\newcommand{\eeqn}{\end{eqnarray}} \newcommand{\bde}{{\bf e}}
\newcommand{\balpha}{{\mbox{\boldmath $\alpha$}}}
\newcommand{\bsalpha}{{\mbox{\boldmath $\scriptstyle\alpha$}}}
\newcommand{\betabf}{{\mbox{\boldmath $\beta$}}}
\newcommand{\bgamma}{{\mbox{\boldmath $\gamma$}}}
\newcommand{\bbeta}{{\mbox{\boldmath $\scriptstyle\beta$}}}
\newcommand{\lambdabf}{{\mbox{\boldmath $\lambda$}}}
\newcommand{\bphi}{{\mbox{\boldmath $\phi$}}}
\newcommand{\bslambda}{{\mbox{\boldmath $\scriptstyle\lambda$}}}
\newcommand{\ggg}{{\boldmath \gamma}} \newcommand{\ddd}{{\boldmath
\delta}} \newcommand{\mmm}{{\boldmath \mu}}
\newcommand{\nnn}{{\boldmath \nu}}
\newcommand{\diag}{{\rm diag}}
\newcommand{\bra}[1]{\langle {#1}|}
\newcommand{\ket}[1]{|{#1}\rangle}
\newcommand{\sn}{{\rm sn}}
\newcommand{\cn}{{\rm cn}}
\newcommand{\dn}{{\rm dn}}
\newcommand{\tA}{{\tilde{A}}}
\newcommand{\tphi}{{\tilde\phi}}
\newcommand{\bpartial}{{\bar\partial}}
\newcommand{\br}{{{\bf r}}}
\newcommand{\bx}{{{\bf x}}}
\newcommand{\bk}{{{\bf k}}}
\newcommand{\bq}{{{\bf q}}}
\newcommand{\bQ}{{{\bf Q}}}
\newcommand{\bp}{{{\bf p}}}
\newcommand{\bP}{{{\bf P}}}
\newcommand{\thet}{{{\theta}}}
\renewcommand{\thefootnote}{\fnsymbol{footnote}}
\

\vskip 0cm
\centerline{
\bf 
Noncommutative Field Theories and Smooth Commutative 
Limits\footnote{This work 
is supported
in part by KOSEF 1998 Interdisciplinary Research Grant
98-07-02-07-01-5 and by BK21 Project of
Ministry of Education.
}}

\vskip .2cm

\vskip 1.2cm
\centerline{ Dongsu Bak,${}^a\!\!$
\footnote{Electronic Mail: dsbak@mach.uos.ac.kr}
 Sung Ku Kim,${}^b$
Kwang-Sup Soh${}^c$ and Jae Hyung Yee${}^d$
}
\vskip 7mm
\centerline{Physics Department, 
University of Seoul, Seoul 130-743 Korea${}^a$}
\vskip0.3cm
\centerline{Physics Department, Ewha Women's University,
Seoul 120-750 Korea${}^b$}
\vskip0.3cm
\centerline{Physics Department, Seoul National University,
Seoul 151-742 Korea${}^c$}
\vskip0.3cm
\centerline{Institute of Physics and Applied Physics, 
Yonsei University,
Seoul 120-749 Korea${}^d$}
\vskip0.4cm
\vskip 3mm

\vskip 1.2cm
\begin{quote}
{
We consider  two model field theories
on a noncommutative plane
that have smooth commutative limits.
One is the single-component fermion theory 
with quartic interaction that vanishes identically 
in the commutative limit.
The other is a scalar-fermion theory, which extends
the scalar field theory with quartic interaction by adding a
fermion.
We compute the bound state energies and the two particle scattering
amplitudes
exactly. 
}
\end{quote}


\newpage

Quantum field theories on noncommutative space 
have recently been studied in both the 
classical\cite{seiberg,connes,douglas,itzhaki,nekrasov,maldacena,hashimoto}
 and quantum
mechanical context\cite{minwalla,filk,gomis,matusis}.
It is one of the most pressing questions whether
the noncommutative quantum field theories are well defined.
In this regards, the nonrelativistic scalar field theory
with a quartic interaction, recently presented in 
Ref.~\cite{yee}, may provide an interesting laboratory
for understanding the nature of the noncommutative 
field theory.
One of the interesting features of the model is that the wave 
functions 
exhibit  the dipole  separation transverse to the total momentum 
in their relative coordinates and that 
the UV/IR mixing does not 
interfere with the renormalizability (at least in the two 
particle sector) of the theory. 

In this note we shall present two model field theories whose
exact wave functions can be constructed and have smooth
commutative limits.
The first model consists of (2+1)-dimensional 
single component fermion field with 
quartic interaction. Although the theory is nontrivial on the 
noncommutative plane, it becomes free in the commutative limit
due to the Pauli exclusion principle.
The other model  consists of the coupled nonrelativistic scalar
and fermionic fields with quartic interaction terms, which has 
analytic but nontrivial commutative limit. This theory is 
the unique extension  of the scalar field
theory with a quartic interaction with the following
restrictions. One is just add a fermion and the other
is that the resulting theory
has its smooth 
commutative limit.   
As in the case of the scalar theory, 
the wave functions of these models have
two center positions in the relative coordinates, 
whose separations grow with the noncommutative scale 
and the total momentum.

{\sl Single Component Fermion Model}

We consider a nonrelativistic single component fermionic 
field theory on a noncommutative plane described by the 
Lagrangian,
\begin{equation}
L=\int d^2x\left( i\,\chi^\dagger \dot\chi +
{1\over 2} \chi^\dagger  \nabla^2
 \psi 
 -{v\over 4} \chi^\dagger * \chi^\dagger * \chi * \chi
\right),
\label{flag}
\end{equation} 
where the  Moyal product ($*$-product) is defined by
\begin{equation}
a(\bx)* b(\bx)\equiv \Bigl(e^{{i}\partial\wedge 
\partial'} a(\bx) b(\bx')\Bigr){\Big\vert}_{\bx=\bx'}
\label{fstar}
\end{equation}
with $\partial\wedge 
\partial'\equiv {1\over 2}\theta\epsilon^{ij} \partial_i \partial'_j$.
In the ordinary commutative spacetime, 
the quantum interaction vanishes due to the Pauli exclusion 
principle 
and the theory becomes free. In the noncommutative plane, however,
it is nontrivial due to the noncommutativity of the space.
As in the case of the scalar field theory\cite{yee},
the scale invariance and the boost part of
the Galilean symmetry are broken
by the explicit scale dependence of the Moyal product. 
The global U(1) invariance under $\chi\!\rightarrow\! e^{i\alpha}
\chi$ persists in the noncommutative case and the number operator
$N=\int d^2x \chi^\dagger\chi$ is conserved enabling one to study 
the system in each N-particle sector separately.
One can quantize the system (\ref{flag}) by using the same procedure
as that of the scalar field theory of Ref.~\cite{yee}. Namely, we shall
adopt the canonical quantization scheme instead of
the path integral quantization in order to incorporate
the normal ordering of the Hamiltonian operator,
\begin{eqnarray}
H=\int d^2x\left(
-{1\over 2} \chi^\dagger  \nabla^2
 \chi 
 +{v\over 4} \chi^\dagger * \chi^\dagger * \chi * \chi
\right)
\label{fhamiltonian}
\end{eqnarray} 
with the canonical anti-commutation relation,
\begin{eqnarray}
\{\chi(\bx),\chi^\dagger(\bx')\}=\delta(\bx-\bx')\,.
\label{fetcr}
\end{eqnarray}
Since the Hamiltonian (\ref{fhamiltonian})
is normal ordered, the two-point Green function
does not receive any perturbative corrections as 
shown in Ref.~\cite{yee} for the bosonic case. 
Thus the full two-point Green function is the same as the 
free propagator,
\begin{eqnarray}
\langle 0|T(\chi(\bx,t)
\chi^\dagger(\bx',t'))|0\rangle 
= \int {d^2 k d\omega\over (2\pi)^3} 
{i\over \omega -{k^2\over 2}+i\epsilon} e^{-i\omega 
(t-t')+i\bk\cdot(\bx-\bx')}
\,.
\label{ftwopoint}
\end{eqnarray} 
Writing the action in the momentum space, one finds the Feynman 
rule for the propagator,
\begin{eqnarray}
S_0(k_0,\bk)=
{i\over k_0 -{k^2\over 2}+i\epsilon} \,,
\label{fpropagator}
\end{eqnarray} 
and for the vertex,
\begin{eqnarray}
\Gamma_0\left(\bk_1,\bk_2;
\bk_3,\bk_4\right)=
iv_0\sin(k_1 \wedge k_2) 
\sin(k_3\wedge k_4)\,, 
\label{fvertex}
\end{eqnarray} 
where $k\!\wedge\! k'={\theta\over 2}\epsilon^{ij}k_i k'_j$.
Note that the vertex vanishes as the commutative parameter $\theta$ 
approaches zero, exhibiting the fermionic nature 
of the theory. 
The propagator and vertex are shown diagrammatically 
in Fig.~1. 
\begin{figure}[tb]
\epsfxsize=2.7in
\hspace{.8in}\epsffile{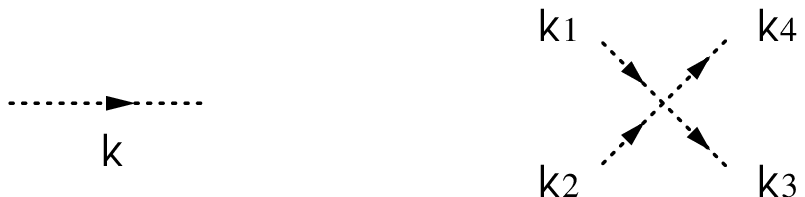}
\\
{\small Figure~1:~The fermion 
 propagator and the vertex.}
\end{figure}
We now compute the on-shell four-point function to obtain the two 
particle scattering amplitude.
The one loop bubble diagram and the full 4-point function are 
depicted in Fig.~2.
\begin{figure}[b]
\epsfxsize=4.5in
\hspace{.8in}\epsffile{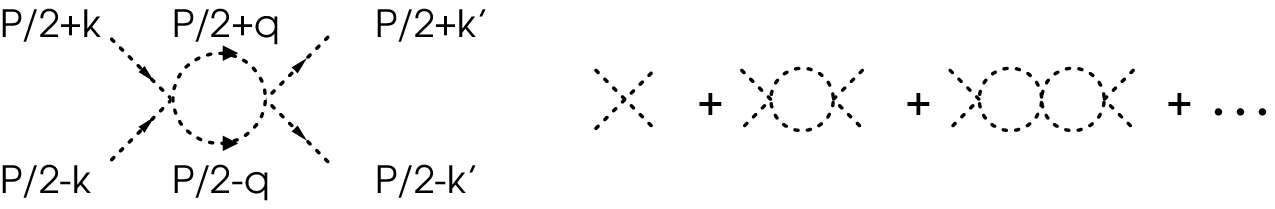}
\\
{\small Figure~2:~The bubble diagram and 
the full  4-point function.}
\end{figure}
Using the Feynman rules one obtains the contribution from the 
one-loop bubble diagram:
\begin{eqnarray}
\Gamma_{\rm b}={1\over 2}(i v_0)^2 i^2 (-1)\int {d^3 q \over (2\pi)^3}\,
{\sin{(P\!\wedge\! k)}\, \sin^2{(P\!\wedge\! q)} \, 
\sin{(P\!\wedge\! k')}
\over \left({W_0\over 2}\! +\!q_0 -{1\over 2}\left({P\over 2}\!+\!q
\right)^2 
\!+\!i\epsilon\right)
 \left({W_0\over 2}\! -\!q_0 -{1\over 2}\left({P\over 2}\!-\!q\right)^2 
\!+\!i\epsilon\right)}
\label{foneloop}
\end{eqnarray}
where $W_0$ is the total energy of the incoming particles,
and the factor $1/2$ and $(-1)$
come from the symmetry of the diagram and the 
fermion loop, respectively.
Introducing the large momentum cut-off $\Lambda$, we find 
the bubble diagram contribution,
\begin{eqnarray}
\Gamma_{\rm b}&=&{-i v_0^2\over 8\pi} \,
\sin(k\wedge P) 
\sin(k'\wedge P)
\int {d^2 q\over 2\pi}{1-\cos(2 q\wedge P)\over
q^2-k^2-i\epsilon}\nonumber\\
&=& {-i v_0^2\over 8\pi} \,
\sin(k\wedge P) 
\sin(k'\wedge P)
 \left[Z\left({\Lambda/k}\right)- 
K_0 (-i\theta k P)\right]
\label{foneloopa} 
\end{eqnarray}
where $K_0(x)$ is the modified Bessel function 
and $Z(x)\equiv  \ln{x}+{i\pi\over 2}$.
It is interesting to compare 
the result (\ref{foneloop}) with that of the 
scalar theory\cite{yee}.
The only difference between  
Eq.~(\ref{foneloop}) and  that of the scalar field theory
is the replacement of the cosine function by the sine function
and the overall negative sign,
which come from the fermionic nature of our system.

Summing all the bubble diagrams of Fig.~2, one obtains 
the full 4-point function,
\begin{eqnarray}
\Gamma ={-i \sin(k\wedge P) 
\sin(k'\wedge P)
 \over {1\over v_0}+{1\over 8\pi}
\left[ Z(\Lambda/k)- 
K_0 (-i\theta k P)\right]}\,.
\label{ffullfour} 
\end{eqnarray}
This result also shows that, in the commutative limit, 
$\theta\!\rightarrow\! 0$,
the 4-point function vanishes and the theory becomes free. Thus in this
fermionic theory,
the noncommutativity of the limits, $\theta\!\rightarrow\! 0$
and $\Lambda\!\rightarrow\! \infty$, does not occur.

One can remove the divergence appearing in Eq.~(\ref{ffullfour})
by redefining the coupling constant as
\begin{eqnarray}
{1\over v(\mu)}= {1\over v_0} + 
{1\over 8\pi}\ln\left({\Lambda\over \mu}\right)\,,
\label{frenormalization} 
\end{eqnarray}
where $\mu$ is a renormalization scale.
For the consistency of this relation, the bare coupling constant
should be negative because $\Lambda$ is supposedly much larger than
the renormalization scale $\mu$. 
The beta function is given by
$\beta(v(\mu))\equiv \mu{\partial v(\mu)\over \partial 
\mu }={v^2(\mu)\over 8\pi}$, which is the same as in the
scalar field theory. 
 
One then obtains the two particle scattering
amplitude,
\begin{eqnarray}
A(\bk,\bk';\bP)={1\over 4\sqrt{\pi k}}\,\,{
\sin(k\wedge P) 
\sin(k'\wedge P)
\over {1\over v(\mu)}+{1\over 8\pi}
[ Z(\mu/k)- 
K_0 (-i\theta k P)]}
\label{famplitude} 
\end{eqnarray}
where an appropriate kinematical factor is included.
The imaginary pole of the scattering amplitude indicates
the existence of the bound states of the two particle system.
Setting the bound state energy $E_B=-\epsilon_B (\epsilon_B >0)$,
it is determined by the solution of the equation
\begin{eqnarray}
{8\pi\over v(\mu)}+\ln\left({\gamma\mu\theta P\over 2}\right)=F(
\sqrt{\epsilon_B}\theta P)\,,
\label{fbound} 
\end{eqnarray}
where $\gamma$ is the Euler constant and $F(x)\equiv \ln (x\gamma/2) + K_0(x)$.
Note that $F(x)$ is monotonically increasing  and covers
the range $[0,\infty)$ for $x\ge 0$. The  bound state does not exist
when the left hand side of (\ref{fbound}) is negative. 
In other words, the bound states cannot be 
formed if $\mu\theta P$ is below the threshold value of $\mu\theta P$.
This reflects the fermionic nature of the theory that, 
in the $\theta\!\rightarrow\! 0$ limit, the theory becomes free.
 There exists one bound state 
when $\mu\theta P\ge 
{2\over \gamma}\, e^{-{8\pi\over v(\mu)}}$.
If $\mu\theta P$ is slightly bigger than the threshold, i.e.
\begin{eqnarray}
\mu\theta P =
\left(1-{\kappa\over 2}\,\ln\left({\kappa  \gamma^2\over e^2}\right)\right)
\left({2\over \gamma}\right) e^{-{8\pi\over v(\mu)}}
\end{eqnarray}
with $0<\kappa\ll 1$,
 the expression for the bound state energy is given by
\begin{eqnarray}
E_B=-{\kappa\, {\gamma^2\mu^2}}\, e^{{16\pi\over v(\mu)}}\,,
\label{fboundenergya} 
\end{eqnarray}
where we have ignored the higher order terms in $\kappa$.
For large $\mu \theta P$, on the other hand, the bound state energy is
given by $E_B=-{\mu^2}e^{{16\pi\over v(\mu)}}$, whose expression
is same as the case of the noncommutative scalar theory.

{\sl Two Particle Schr\"odinger Equation}

Since the number operator of the system (\ref{flag}) is conserved, one may 
treat the system quantum mechanically by constructing the Schr\"odinger
equation in each N-particle sector. The Schr\"odinger equation for the two 
particle wave function, 
$\Psi(\br,\br')\!=\!\langle 0|\chi(\br,t)\chi(\br',t)|\Phi\rangle$ can be 
obtained from the operator Schr\"odinger equation,
\begin{eqnarray}
i\dot\chi=[\chi, H]
=-{1\over 2}\nabla^2 \chi+ {v\over 4}
\int d^2x\left(  \delta(\br-\bx)* \chi^\dagger- \chi^\dagger* 
\delta(\br-\bx)\right) 
* \chi* \chi\,.
\label{fopeq}
\end{eqnarray} 
The two particle Schr\"odinger equation reads
\begin{eqnarray}
i\dot{\Psi}(\br,\br')\!=\!-
{1\over 2}\!\left(\nabla^2\!+\!{\nabla'}^2\right)
\Psi(\br,\br')\!+\! {v\over 4}
\int d^2x\left[  \delta(\br'\!-\!\bx)\!*\!  
\delta(\br\!-\!\bx)\!-\! 
\delta(\br\!-\!\bx)\!*\! 
\delta(\br'\!-\!\bx)\right]\!*\! 
\Psi_*(\bx,\!\bx)\,,
\label{ftwobody}
\end{eqnarray}  
where $\Psi_*(\bx,\!\bx')=e^{i\partial\wedge
\partial'} \Phi(\bx,\bx')$.
In the momentum space the Schr\"odinger equation becomes
\begin{eqnarray}
i\dot{\Psi}(\bQ,\bq)\!=\left({1\over 4}{Q^2}+q^2\right)
\Psi(\bQ,\bq)\!+\! 
{v_0 \sin(q\wedge Q)\over 8\pi^2}
\int d^2 q' \sin(q'\wedge Q)
\Psi(\bQ,\bq')\,,
\label{ftwobodyaa}
\end{eqnarray}  
where $Q$ and $q$ are the total and the relative momenta
of the two particles, respectively.
Here again, the equation (\ref{ftwobodyaa}) for the fermion system 
can be obtained by replacing the cosine function by sine function
in the equation for the scalar theory.

Writing 
\begin{eqnarray}
\Psi(\bQ,\bq)= \delta(\bQ\!-\!\bP)\tilde\chi(\bP,\bq) 
e^{-i({1\over 4}{P^2}+E_r)t}\,,
\end{eqnarray}  
the momentum space
 equation is reduced to 
\begin{eqnarray}
(E_r-q^2)\tilde\chi(\bP,\bq)=
-\sin(q\wedge P)C(\bP)\,,
\label{ftwobodyb}
\end{eqnarray}  
with $C(\bP)=
-{v_0\over 8\pi^2}\int d^2 q' \sin(q'\wedge P)
\tilde\chi(\bP,\bq')$.
For the bound state, we obtain 
\begin{eqnarray}
\tilde\chi(\bP,\bq)=
{ \sin(q\wedge P)\over q^2+ \epsilon_B} C(\bP)\,.
\label{frelative}
\end{eqnarray}  
with $E_r=-\epsilon_B$.
 Integrating both sides with weighting factor
$\sin( q\wedge P)$, one obtains the 
eigenvalue equation (\ref{fbound})
for the bound state after renormalization.
Thus the result of the two particle Schr\"odinger equation agrees
with that of the perturbative theory.
The bound state wave function in the position space can be obtained 
from (\ref{frelative}):
\begin{eqnarray}
\Psi =e^{i{\bP}\cdot {\bf R}}
\left[K_0\left(\sqrt{\epsilon_B}\, |\bx+{\!\!\!\phantom{1}_1\over 
\!\!\!\phantom{1}^2}\theta \tilde\bP|\right)- 
K_0\left(\sqrt{\epsilon_B}\, 
|\bx-{\!\!\!\phantom{1}_1\over 
\!\!\!\phantom{1}^2}\theta \tilde\bP|\right)\right]\,,
\label{boundenergya}
\end{eqnarray} 
where $\tilde{P}^i=\epsilon^{ij}P_j$, ${\bf R}={\br_1\!+\!\br_2\over 2}$
and the relative position $\bx$ denotes $\br_1\!-\!\br_2$.
As in the scalar field theory the wave function has two centers
at   $\pm {1\over 2}\theta \tilde\bP$
in the relative coordinates. 
The separation of these centers grows linearly in $\theta P$
and is perpendicular to the total momentum $\bP$, exhibiting
explicitly the stringy dipole nature of the theory\cite{susskind}.

The scattering solution can also be found with $E_r=k^2$:
\begin{eqnarray}
\tilde\chi(\bP,\bq)=(2\pi)^2\delta(\bq-\bk)
+C(\bP) {\sin (q\wedge P)\over q^2-k^2-i\epsilon} \,,
\label{scatteringsol}
\end{eqnarray}     
where $C(\bP)$ is as given above.
Proceeding similarly as in the case of the scalar field theory, one 
obtains the exact wave function for the two particle scattering: 
\begin{eqnarray}
\tilde\chi(\br)=e^{i\bk\cdot \br}+ {1\over 8}
\left[H^{(1)}_0\left(k |\br+{\!\!\!\phantom{1}_1\over 
\!\!\!\phantom{1}^2}\theta \tilde\bP|\right)-
H^{(1)}_0\left(k |\br- 
{\!\!\!\phantom{1}_1\over 
\!\!\!\phantom{1}^2}\theta \tilde\bP|\right) \right]C(\bP)\,.
\label{scatteringsolb}
\end{eqnarray} 
where $H^{(1)}_\nu (x)$ is the Hankel function of the  first kind. 
Once $C(\bP)$ is determined self consistently, one may check that 
the resulting scattering amplitude agrees precisely with
that of the perturbation theory. The scattering solution again has two 
dipole centers 
whose separation grows linearly with $\theta P$
in the perpendicular direction of $\bP$.

{\sl Coupled Scalar-Fermion System}

As another nontrivial example of the noncommutative 
field theory that has the smooth
commutative limit, we consider the coupled scalar-fermion theory
described by the Lagrangian
\begin{equation}
L=\!\int d^2x\left( i\,\psi^\dagger \dot\psi\! +\! i\,\chi^\dagger \dot\chi
\!-\!{1\over 2}| \nabla\psi|^2\!-\!{1\over 2}| \nabla\chi|^2  
 \!-\!{v\over 4}[\psi^\dagger\! *\! \psi^\dagger \!+\!\chi^\dagger\! *\! \chi^\dagger]
\!*\! \left[\psi\! *\! \psi\!+\! \chi\! *\! \chi\right]
\right),
\label{sflag}
\end{equation} 
where $\psi(\bx,t)$ and $\chi(\bx,t)$ represent
a scalar and a single-component fermion field operators, respectively.
The canonical (anti)-commutation relations among these fields
are given by
\begin{eqnarray}
&&[\psi(\bx),\psi^\dagger(\bx')]=\delta(\bx-\bx'), \ 
[\psi(\bx),\psi(\bx')]=0\nonumber\\
&&\{\chi(\bx),\chi^\dagger(\bx')\}=\delta(\bx-\bx'),\ 
 \{\chi(\bx),\chi(\bx')\}=\{\chi^\dagger(\bx),\chi^\dagger(\bx')\}=0\,,
\label{sfetcr}
\end{eqnarray}
while all the remaining commutators are vanishing.
In the commutative limit, the fermionic interaction terms vanish and thus the theory 
reduces to the ordinary nonrelativistic scalar theory with
quartic interaction\cite{jackiw, bergman}.
The global U(1) invariance under 
$\psi\!\rightarrow\! e^{i\alpha}\psi$ and
$\chi\!\rightarrow\! e^{i\alpha}\chi$ insures the conservation of
total number operator $N=\int d^2 x\,(\psi^\dagger \psi+\chi^\dagger \chi)$.
However, the scalar and fermionic number
operators are respectively  conserved 
only for $\theta=0$. The violation of 
the the fermion number conservation is of 
order $\theta$ as will 
be seen below and will disappear 
in the commutative limit. There is an additional $Z_2$
symmetry that exchanges the fermion and boson. It is this 
symmetry that is  partly responsible for  the smooth
commutative limit of the system. 

Feynman rules for the perturbative calculations can be 
obtained as 
in the fermionic theory. The scalar and fermion 
propagators are the same 
as the ones in Ref.~\cite{yee} and Eq.~(\ref{fpropagator}),
respectively. There are now 
four kinds of vertices.   
One can evaluate the Green function and study
the bound state and the scattering problem perturbatively
as done in the earlier cases. Instead we shall use the quantum 
mechanical approach 
to compute the exact wave functions for the sector
of two bosonic and two fermionic particles. The coupled 
Schr\"odinger equation for the two boson state 
$\phi(\br,\br')\!=\!\langle 0|\psi(\br,t)\psi(\br',t)|\Phi\rangle$
and two fermion state
$\Psi(\br,\br')\!=\!\langle 0|\chi(\br,t)\chi(\br',t)|\Phi\rangle$,
can be constructed from the operator equation of motion for
the field operator $\psi$ and $\chi$.
In the momentum space they read
\begin{eqnarray}
&&i\dot{\phi}(\bQ,\bq)\!=\left({1\over 4}{Q^2}+q^2\right)
\phi(\bQ,\bq)\nonumber\\
&&\ \ \ \ + 
{v_0 \cos(Q\wedge q)\over 8\pi^2}
\int d^2 q'\left(\cos(Q\wedge q')
\phi(\bQ,\bq')\!+\! i\sin(Q\wedge q')
\Psi(\bQ,\bq')\right)\,,
\label{sftwobodyaa}
\end{eqnarray}  
and 
\begin{eqnarray}
&&i\dot{\Psi}(\bQ,\bq)\!=\left({1\over 4}{Q^2}+q^2\right)
\Psi(\bQ,\bq)\nonumber\\
&&\ \ \  \ - 
{iv_0 \sin(Q\wedge q)\over 8\pi^2}
\int d^2 q'\left(\cos(Q\wedge q')
\phi(\bQ,\bq')\!+\! i\sin(Q\wedge q')
\Psi(\bQ,\bq')\right)
\,,
\label{sftwobodybb}
\end{eqnarray}  
where $Q$ and $q$ are the total and the relative momenta
of the two particles, respectively.
Setting 
\begin{eqnarray}
&&\phi(\bQ,\bq)= \delta(\bQ\!-\!\bP)\varphi(\bP,\bq) 
e^{-i({1\over 4}{P^2}+E_r)t}\nonumber\\
&&\Psi(\bQ,\bq)= \delta(\bQ\!-\!\bP)\tilde\chi(\bP,\bq) 
e^{-i({1\over 4}{P^2}+E_r)t}
\,,
\label{sfoverall}
\end{eqnarray}  
we finally obtain the coupled Schr\"odinger equation for the
relative motion:
\begin{eqnarray}
&& E_r{\varphi}(\bP,\bq)\!=q^2
\varphi(\bP,\bq) -\cos(Q\wedge q)G(\bP)
\nonumber\\
&& E_r{\tilde\chi}(\bQ,\bq)\!=q^2
\tilde\chi(\bP,\bq) +i\sin(Q\wedge q)G(\bP)
\,,
\label{sfrelative}
\end{eqnarray}  
with
\begin{eqnarray}
G(\bP)=-{v_0 \over 8\pi^2}
\int d^2 q'[ \,\cos(q'\wedge Q)
\varphi(\bP,\bq')\!+\! i\sin(q'\wedge Q)\tilde\chi(\bP,\bq')]
\,.
\label{sfg}
\end{eqnarray} 

The bound state energy is obtained by setting $E_r=-\epsilon_B$.
Then the Schr\"odinger equation implies that
\begin{eqnarray}
\varphi(\bP,\bq)=
{\cos(k\wedge Q)\over k^2 +\epsilon_B} G(\bP),\ \ 
\tilde\chi(\bP,\bq)=
-i{\sin(k\wedge Q)\over k^2 +\epsilon_B} G(\bP)
\,.
\label{sfbound}
\end{eqnarray}  
Substituting (\ref{sfbound}) into (\ref{sfg}), one finds the 
eigenvalue equation for the bound state energy,
\begin{eqnarray}
1=-{v_0\over 8\pi^2}\int d^2 q {1\over  q^2 +\epsilon_B}= 
=-{v_0\over 4\pi} 
\ln\left({\Lambda\over \sqrt{\epsilon_B}}\right)\,.
\label{sfboundaa}
\end{eqnarray}  
By redefining the bare coupling constant as
\begin{eqnarray}
{1\over v(\mu)}= {1\over v_0} + 
{1\over 4\pi}\ln\left({\Lambda\over \mu}\right)\,,
\label{sfrenormalization} 
\end{eqnarray}
where $\mu$ is the renormalization scale, one obtains the bound
state energy,
\begin{eqnarray}
E_r= -\mu^2 e^{8\pi\over v(\mu)}
\,.
\label{sfenergy} 
\end{eqnarray}
This bound state energy is exactly the same as
that of the ordinary scalar theory with
the contact interaction\cite{jackiw,bergman}.
The reason for this is that the nonplanar parts of fermionic 
loop contribution exactly cancel those of the bosonic loops, 
and that only the planar diagrams contribute to 
the 4-point Green functions. Thus the theory has the 
analytic limit as $\theta\!\rightarrow\! 0$.
The beta function, $\beta(v)={v^2\over 4\pi}$ 
also agrees with 
the commutative scalar theory\cite{jackiw,bergman}
for the same reason.

The scattering problem can also be dealt with
as in the earlier cases.
We consider scattering state of two incident 
scalar particles
with relative energy $E_r=k^2$.
The solution of the Schr\"odinger 
equation (\ref{sfrelative}) reads
\begin{eqnarray}
&&\varphi(\bP,\bq)=(2\pi)^2\delta(\bq-\bk)
+G(\bP) {\cos (q\wedge P)\over q^2-k^2-i\epsilon} \,, \nonumber\\
&&
\tilde\chi(\bP,\bq)=
-iG(\bP) {\sin (q\wedge P)\over q^2-k^2-i\epsilon} 
\label{sfscatteringsol}
\end{eqnarray}     
where $G(\bP)$ is as given above.
Substituting (\ref{sfscatteringsol})
to (\ref{sfg}), one finds
\begin{eqnarray}
G(\bP) =-{1\over 2}\,\,{\cos (q\wedge P)\over {1\over v_0}+
{1\over 4\pi} Z(\Lambda/k)}\,.
\label{sfgg}
\end{eqnarray}     
Thus the exact wave function in coordinate space
becomes
\begin{eqnarray}
&&\varphi(\br)=e^{i\bk\cdot \br}+ {i\over 8}
\left[H^{(1)}_0\left(k |\br+{\!\!\!\phantom{1}_1\over 
\!\!\!\phantom{1}^2}\theta \tilde\bP|\right)+
H^{(1)}_0\left(k |\br- 
{\!\!\!\phantom{1}_1\over 
\!\!\!\phantom{1}^2}\theta \tilde\bP|\right) \right]G(\bP)\nonumber\\
&&\tilde\chi(\br)={1\over 8}
\left[H^{(1)}_0\left(k |\br+{\!\!\!\phantom{1}_1\over 
\!\!\!\phantom{1}^2}\theta \tilde\bP|\right)-
H^{(1)}_0\left(k |\br- 
{\!\!\!\phantom{1}_1\over 
\!\!\!\phantom{1}^2}\theta \tilde\bP|\right) \right]G(\bP)
\label{sfscatteringsolb}
\end{eqnarray} 
This result again exhibits the dipole nature 
in the relative
coordinate having two centers.

In the commutative limit, $\theta\!\rightarrow\! 0$, the bosonic wave 
function reduces to that of the ordinary scalar theory of 
Ref.~\cite{jackiw}, and the fermion wave function vanishes. 
Thus this results clearly indicate that the coupled
scalar fermion theory in (\ref{sflag})
has analytic commutative limit.
By taking asymptotic limit of (\ref{sfscatteringsolb}),
one finds the scattering amplitudes; For the outgoing 
two bosons,
\begin{eqnarray}
A_b(\bk,\bk';\bP)=-{1\over 4\sqrt{\pi k}}\,\,{
\cos(k\wedge P) 
\cos(k'\wedge P)
\over {1\over v(\mu)}+{1\over 4\pi}
 Z(\mu/k)}
\label{bamplitude} 
\end{eqnarray}
and for the out going two fermions,
\begin{eqnarray}
A_f(\bk,\bk';\bP)=-{1\over 4\sqrt{\pi k}}\,\,{
\cos(k\wedge P) 
\sin(k'\wedge P)
\over {1\over v(\mu)}+{1\over 4\pi}
 Z(\mu/k)}\,,
\label{sfamplitude} 
\end{eqnarray}
where $k$ and $k'$ are respectively the initial and the final 
relative momenta.
The fermion production out of two boson scattering 
is zero when the final relative 
momentum is parallel to the the total momentum and has a 
peak when the final relative momentum is perpendicular to
the total momentum. Of course, the production vanishes 
when $\theta=0$, so the limit $\theta\!\rightarrow\! 0$
is analytic. 

In this note, we have presented
two noncommutative  field theories 
that have smooth commutative limits. The single component fermion 
model is intriguing because
the quartic interaction was identically vanishing
on the ordinary plane but produces nontrivial
interactions due to the noncommutativity of the geometry.

In the noncommutative 
scalar theory with a quartic interaction of Ref.~\cite{yee},
sending the momentum cut-off to infinity does not
commute with the operation $\theta P\!\rightarrow\! 0$ though this
does not interfere with the renormalizability 
of the theory. In other words, the $P\theta=0$ theory
differs from the theory defined by taking 
$P\theta\!\rightarrow\! 0$ limit  of
the regulated  theory with a nonzero $\theta P$.
Although not conclusive, this discontinuity might even spoil
the renormalizability  in the case of the relativistic 
quantum field theories. The only nontrivial exception so far found 
seems the noncommutative 
N=4 Super Yang-Mills theory where 
the UV/IR mixing does not occurs\cite{matusis}.

We here asked whether there exists a theory free of the
discontinuity at $\theta P=0$ which extends the scalar theory with 
a quartic interaction by adding a fermion.
The supersymmetric extension of the scalar
Schr\"odinger theory
is not the desired one because 
it may be 
verified that there are still discontinuities although 
the theory is renormalizable. The answer is rather 
unique and it is nothing 
but the coupled scalar-fermion theory presented  
here. 
One consequence of the continuity requirement is that,
in the coupled theory, the fermion number conservation
is violated  by order of $\theta$.

\vfill
\eject


\begin{thebibliography}{99}

\bibitem{seiberg}
N. Seiberg and E. Witten, 
JHEP (1999) 9909:032, 
  {\tt hep-th/9908142}. 

\bibitem{connes}
A. Connes, M.R. Douglas, and A. Schwarz, 
JHEP {\bf 02} (1998) 003, {\tt hep-th/9711162}.

\bibitem{douglas}
M.R. Douglas and C. Hull, 
  JHEP {\bf 02} (1998) 008, {\tt hep-th/9711165}.


\bibitem{itzhaki}
A.~Hashimoto and N.~Itzhaki, 
Phys. Lett. {\bf B465} (1999) 142, {\tt
  hep-th/9907166}.


\bibitem{nekrasov}
N.~Nekrasov and A.~Schwarz, 
Commn. Math. Phys. {\bf 198} (1998) 689,  {\tt
  hep-th/9802068}.

\bibitem{maldacena}
J.~M. Maldacena and J.~G. Russo, 
JHEP(1999) 9909:025,  
{{\tt hep-th/9908134}}.



\bibitem{hashimoto} A. Hashimoto and K. Hashimoto, 
JHEP 9911 (1999) 005,
{\tt hep-th/9909202}; D. Bak, Phys. Lett. {B 471} (1999) 149,
{\tt hep-th/9910135}.


\bibitem{minwalla}
S. Minwalla, M. V. Raamsdonk and N. Seiberg, {\tt hep-th/9912072}. 





\bibitem{filk}
T. Filk, 
Phys. Lett. {\bf B376} (1996) 53;
%
M. Chaichian, A. Demichev, P. Presnajder, 
Nucl. Phys. {\bf B567} (2000) 360, {\tt hep-th/9812180}; 
%
H. Grosse, T. Krajewski, R. Wulkenhaar,
{\tt hep-th/0001182}; 
%
M. Hayakawa, Phys. Lett. {\bf B478} (2000) 394,
{\tt hep-th/9912094}. 

\bibitem{gomis}
J. Gomis, K. Landsteiner and E. Lopez, {\tt hep-th/0004115}. 

\bibitem{matusis}
A. Matusis, L. Susskind and  N. Toumbas, {\tt hep-th/0002075}. 


\bibitem{yee}
D. Bak, S. K. Kim, K.-S. Soh and J. H. Yee,
{\tt hep-th/0005253}. 



\bibitem{jackiw}
R. Jackiw, 
{\sl Delta function potential in two and three 
dimensional quantum mechanics} in Beg Memorial Volume, 
A. Ali and P. Hoodbhoy eds. (World Scientific, Sigapore,1991).


\bibitem{bergman}
O. Bergman, Phys. Rev. {\bf D46} (1992) 5474. 



\bibitem{susskind}
D. Bigatti and L. Susskind, 
{\tt hep-th/9908056}.

  













\end{thebibliography}
\end{document}